# Protection of Cities from Small Rockets, Missiles, Projectiles and Mortar Shells

**Alexander Bolonkin**
C&R, 1310 Avenue R, #F-6, Brooklyn, NY 11229, USA
T/F 718-339-4563, aBolonkin@juno.com, http://Bolonkin.narod.ru
**Joseph Friedlander**
Shave Shomron, Israel 44858, jjfriedlander@hotmail.com

## Abstract

The authors suggest a low cost closed AB-Dome, which may protect small cities such as Sederot from rockets, mortar shells, chemical and biological weapons. The offered AB-Dome is also very useful in peacetime because it protects the city from outside weather (violent storms, hail) and creates a fine climate within the Dome. The roughly hemispherical AB-Dome is a gigantic inflated thin transparent film, located at altitude up to 1 – 5 kilometers, which converts the city into a closed-loop air system. The film may be armored with a basalt or steel grille or cloth pocket-retained stones that destroy (by collision or detonation) incoming rockets, shells and other projectiles. Such an AB-Dome would even protect the city in case of a third-party nuclear war involving temporary poisoning of the Earth's atmosphere by radioactive dust. The building of the offered dome is easy; the film spreads on the ground, the fan engines turn on and the cover rises to the needed altitude and is supported there by a small internal overpressure.

The offered method is cheaper by thousands of times than protection of a city by current anti-rocket systems. The AB-Dome may be also used (height is up to 1-5 and more kilometers) for TV, communication, long distance location, tourism, suspended high speed and altitude windmills (energy), illumination and entertainment (projected messages and pictures).

The authors developed the theory of AB-Dome, made estimations, computations and computed a typical project. Discussion and results are at the end of the article.

Note: Some background material in this article is gathered from Wikipedia under the Creative Commons license.

## Introduction

One important problem in small countries with hostile borders (or larger countries with leaky borders) is protection of domestic targets from small rockets, missiles and mortar shells. For well over a hundred years there has been no satisfactory solution for this, which is why such weapons are favorites of guerilla groups. Israel, for example, has villages (Alumim, and dozens of others) towns (Sderot, Netivot and Qiryat Shemona), cities such as Ashqelon and numerous installations near unfriendly borders. For example, Sderot lies one kilometer from the Gaza Strip and the town of Beit Hanoun. Since the beginning of the Second Intifada in October 2000, Sederot has been under constant rocket fire from Qassam rockets launched by various armed factions. Despite the imperfect aim of these homemade projectiles, they have caused deaths and injuries, as well as significant damage to homes and property, psychological distress and emigration from the city. Real estate values have fallen by about half. The Israeli government has installed a "Red Dawn" alarm system to warn citizens of impending rocket attacks, although its effectiveness has been questioned. Thousands of Qassam rockets have been launched since Israel's disengagement from the Gaza Strip in September 2005, which essentially has killed popular support for any further withdrawals, particularly from West Bank areas near the heart of the country. Even pro 'land for peace' parties have no plausible answer to the question (in its various forms), "And what will you do when they start shooting at (the) Tel Aviv (stock exchange) and the (Ben Gurion International) airport?" Thus, even with perhaps 70 of 120 votes in the Israeli Parliament (Knesset) in favor of negotiations with the Palestinian Authority, no final 'end of conflict' agreement is seriously in prospect, nor is likely to be given the impossibility of guaranteeing a stoppage of the bombardments by *some* unsatisfied Palestinian faction—even one not yet on the scene.

In May 2007, a significant increase in shelling from Gaza prompted the temporary evacuation of thousands of residents. By November 23, 2007, 6,311 rockets had fallen on the city. The Israeli newspaper Yediot Aharonot reported that during the summer of 2007, 3,000 of the city's 22,000 residents (comprised mostly of the city's key upper and middle class residents, the heart of the economy, those most able to move,) had already left for other areas, out of Qassam rocket range. Russian-Israeli billionaire Arkady Gaydamak has in recent years supported relief programs for residents who cannot leave. [5]

On December 12, 2007, on a day during which more than *20* rockets landed in the Sderot area, including a direct hit to one of the main avenues, the mayor of Sderot, Eli Moyal (a well-known figure in Israeli media) unexpectedly announced his resignation from the job, citing the government's failure to stop the daily rocket attacks. "Maybe this will spark the government to launch an operation for the lives of its [Sderot's] residents. I can't keep making the decisions, they can't keep piling it all on me," Mr. Moyal reportedly said. Later under political pressure, he was asked to resume his job by key national leaders, as a duty rather than a choice.

Qassams were first fired at Israeli civilian targets in October 2001.The first Qassam to land in Israeli territory was launched on February 10, 2002. The first time an Israeli city was hit was on March 5, 2002, when two rockets struck Sderot. Some rockets have hit as far as the edge of Ashkelon. The total number of Qassam rockets launched exceeded 1000 by June 9, 2006. During the year 2006 alone, 1000+ rockets were launched. Tons of explosives have been intercepted at the Egyptian border; the uninterrupted shipments must be greater still, and the cumulative detonation yield has easily been in the tens of tons.

The introduction of the Qassam rocket took Israeli politicians and military experts by surprise. Reactions have been mixed. The Israeli Ministry of Defense views the Qassams as "more a psychological than physical threat." The rockets are fired largely at civilian populations. There is some evidence of psychological damage to children in the effected areas, particularly in Sderot. The IDF has reacted to the deployment of the Qassam rockets by deploying the Red Color early warning system in Sderot, Ashkelon and other at-risk targets. The system consists of an advanced radar that detects rockets as they are being launched. Loudspeakers warn civilians to take cover approximately twenty seconds before impact in an attempt to minimize the threat posed by the rockets.

A rocket once fell into the electricity station in Ashkelon and caused electricity shortages in several areas, other time a rocket-similar to Qassam- fell inside an army base and injured more than 70 Israeli soldiers. The Ashkelon strike in particular was troubling as it added (by its radius) another 250,000 Israelis to the potential target list requiring defenses to be paid for, active or passive.

Some military bases of the USA in Afghanistan and Iraq (or in various parts of Asia and Latin America) are in the same situation. Any security consultant working to protect valuable installations in the more volatile corners of Africa, Latin America or Asia will recognize the dangers in the scenarios listed above. Rockets, and remotely triggered mortars, are man-portable and can be smuggled in, can be covertly emplaced and remotely fired with no appreciable warning, and endanger billions in investment with mere thousands in expenses. In Gaza, bonuses are allegedly given to impoverished children to retrieve the launchers, to reduce the expenses of replacing both rounds and launchers.

**Israel Government plan**.

*(As reported at www.haaretz.com 24/12/07 http://www.haaretz.com/hasen/spages/937756.html), the plan of the Israeli government was to fund an anti-missile system, but not to reinforce the homes of all Sderot-area residents, which caused considerable local anger.*
*Many poorer residents don't have a reinforced secure room they can retire to when the alarm sounds, with metal in the walls, proof against shrapnel.*
*The Government's position was that the promise to armor all the homes was made before it was understood that the anti-rocket system would cost the same 1.5 billion NIS (New Israeli Shekels) that reinforcing all the homes would. Critics object that the anti-rocket system may fail in at least a fraction of the cases to work, at which point those on the ground without a passive defense (a safe room) would be in mortal danger.*
.
*Related articles:*
- *Security cabinet okays funding for 'Iron Dome' rocket defense system*
- *Defense Minister selects Rafael anti-missile defense system*
- *Sderot to keep 'special status' due to Qassam fire*
- *Court: State must brief court on plans to reinforce Sderot homes</A< span>*

*The currently offered RAFAEL anti-rocket system is shown below:*

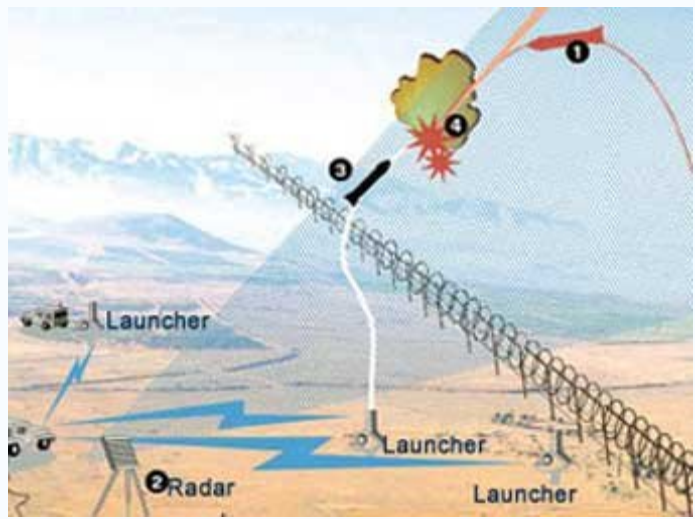

*The RAFAEL anti-rocket system (Iron Dome) costs about 1 billion American dollars. It may be ready in 2011. Efficiency of operational system remains unknown.*

The other system, such as "David's Sling" or C-RAM from Raytheon have perhaps 70 – 80% efficiency. That means every third to fifth missile would reach the target. From an investor's standpoint, this would be little better than unchecked bombardment; capital would still flee the targeted city. What is needed is a defense so thorough that the residents are *entirely unaware* of the bombardment other than possibly distant flashes and evening news summaries. Any difference in daily life that a bombardment causes ultimately limits the ability to conduct business as usual (with no risk premium). Anything less than this standard still makes the inhabitants feel under seige.

**Quassam rockets**.

**Table 1.** Specifications of several types of Qassam (home workshop class) rockets

|  | Qassam 1 | Qassam 2 | Qassam 3 |
|---|---|---|---|
| Length (cm) | 79 | 180 | 299+ |
| Diameter (cm) | 6 | 15 | 17 |
| Weight (kg) | 5.5 | 32 | 90 |
| Explosives Payload kg) | 0.5 | 5 - 7 | 10 |
| Maximum range (km) | 3 | 8-10 | 10 |

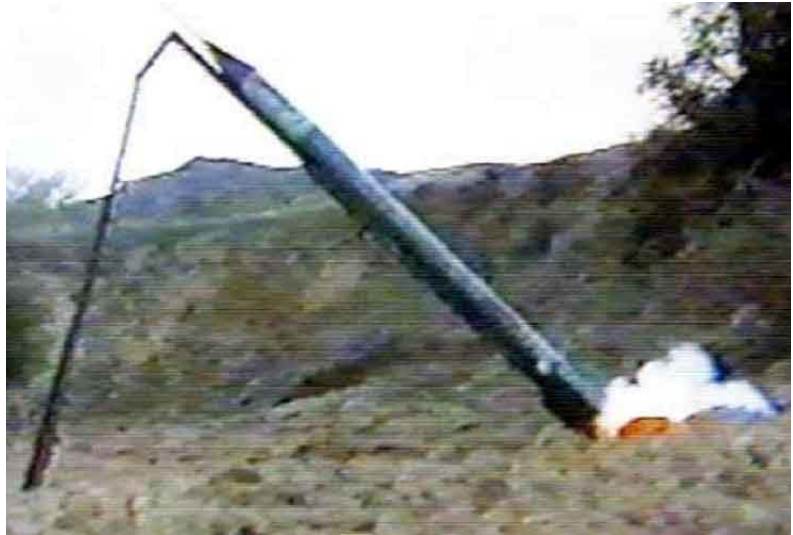
Qassam (Kassam) rocket

The Qassam rockets are tens (hundreds) times cheaper than the complex electronic anti-rockets and their delicate support system. The mortar shells are cheaper by hundreds of times. Attempted defense from them by conventional anti-rocket system may ruin any rich country, and in the end not work anyway, because the enemy always has the option of using large salvos, to probe the point where the system collapses trying to defeat X simultaneous launches.

**Basalt for use as blocking grid**

Basalt is volcanic magma (effusive rock) widely distributed cheap material in the Earth, One melts the temperature about 1500 K (1200 C), may be cast in grille and has good strength. Basalt has an extraordinarily low viscosity which is necessary for superior basalt castings. The mechanical properties of basalt is comparable to those of cast iron and many fine steels, and superior to aluminum, brass, bronze, and copper both in compression and shear strengths. Basalt is good matter for grills because it is very cheap (about $10-20 per ton) and may be cast in the shape of the needed grille. The basalt may be reinforced and secured in place by cheap basalt fiber having a very high tensile strength. A list of the properties of cast basalt is collected in Table 2. (This data is derived from a U.S. Government publication)

**Table 2.- Properties Of Cast Basalt**

| Physical properties | Average numerical value, MKS units |
|---|---|
| Density of solid | 2900-2960 kg/m$^3$ |

| Tensile strength | $3.5 \times 10^7$ N/m$^2$ |
|---|---|
| Compressive strength | $5.4 \times 10^8$ N/m$^2$ |
| Bending strength | $4.5 \times 10^7$ N/m$^2$ |
| Modulus of elasticity (Young's modulus) | $1.1 \times 10^{11}$ N/m$^2$ |
| Moh's hardness | 8.5 |
| Grinding hardness | $2.2 \times 10^5$ m$^2$/m$^3$ |
| Melting point | 1400-1600 K |
| Hygroscopicity | 0.1% |

## Description of AB-Dome and Innovations

The authors offer a new protection system against warheads from kilogram range all the way up to Hiroshima-yield nuclear weapons. That is the AB-Dome as described in works [1]-[4].

The idea is a dome covering a city by a thin transparent film 2 (fig.1). The film has thickness of 0.05 – 0.3 mm. This may be located at high altitude (0.2 – 0.5 kilometers—and if defense against Qassam rockets weapons is desired, the higher figure is definitely desired, and may even be increased). The film is supported at this altitude by a small additional air pressure produced by ground ventilators. That may be connected to Earth's ground by managing cables 5. The film may have a controlled transparency (option). The system may also have a second lower film 6, possibly in the form of a mesh, which protects the city from fragments of rockets (option). Cluster munitions, which may be, depending on the model, the size of soda cans, cell phones, baseballs, or D cell batteries, may have their outer carrier detonated on the outside of the dome and the hundreds of submunitions may roll down the exterior of the dome until their trailing bomblet tail, spinning, unscrews the firing pin and detonates each bomblet, kilometers distant from the city below. Options are listed below, including a mesh, for catching these small fragments.

Even if not caught, the fragments impact at mere terminal velocity, not their initial speed (perhaps 1000-1500 m/s a few meters from the detonation point.)

Terminal velocity on a raindrop of 1.5 mm radius is given at http://www.gantless.com/paper.html as about 7 meters a second. Thus a small iron fragment should sting as Macklam, Janos et.al there conclude, or cut, or worse, but not do terrible damage in most cases to unshielded humans, certainly nothing in comparison to the horrific effect of a shredding cluster bomblet at ground level without an AB-Dome.

(The minimum 3-5% of submunitions that do not detonate would be found around the perimeter of the dome at ground level, concentrated for easy collection and disposal, instead of scattered cunningly, hanging from trees, inside barn windows, in ground crevasses, and other dangerous spots. In addition, there is the distinct possibility that the non-rigid nature of the AB-Dome would deform and rebound from the explosion, leaving much of the shrapnel on the outside. This possibility warrants further modeling.)

Means of repair are available to close the hole at the top armored film resulting from the exploded rocket without exposing the entire city to further attack.

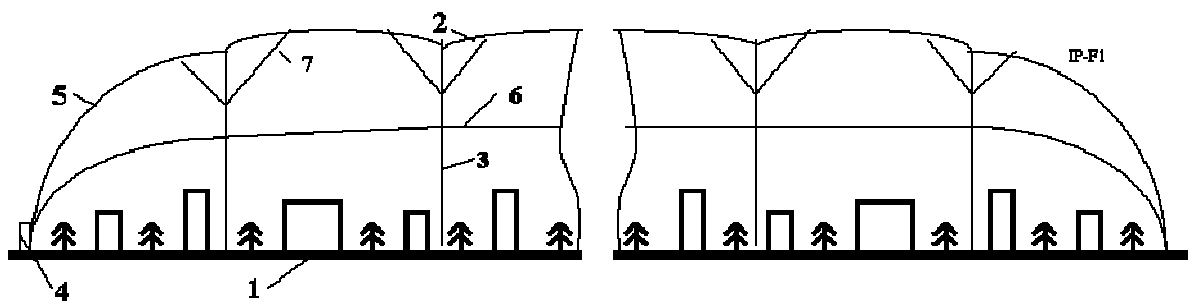

**Fig. 1.** Film AB-Dome for given city. *Notations*: 1 - area, 2 - thin film cover with variable clarity (option), 3 – control support cable length (height) is 0.2 – 5 kilometers), 4 - exits and ventilators, 5 - border section (as opposed to interior section—part of the AB-Dome's periphery, requiring stouter construction), 6 – the second (lower) controlled reflectivity film cover (option), 7 – additional support cables.

Even a small additional overpressure creates a significant lift force. For example, the small overpressure of only only $p = 0.01$ atmosphere produces a lift force of about 100 kg/m$^2$. At altitude of $H = 5$ kilometers, owing to the lesser density of the air at that height, this force is more like 53 kg/m$^2$. The support cable has a weight of about 1-3 kg/m$^2$, the 1 m$^2$ of film weighs less than 0.05 - 0.5 kg (for example, kevlar film of thickness ~0.2 mm has the weight of 0.3 kg/m$^2$). *That means that every square meter of dome can keep at altitude a useful load from 50 - 95 kg*. At high altitude the useful load decreases, but if it is needed, we can increase the overpressure. As the fans are at ground altitude for easy maintenance, this is no technical problem (though an additional operating expense).

The top film may support, for example, a basalt grating 8 (fig.2) on the upper surface. If the grille has a mesh of rod 1×2.5 cm of cross-section and a step (grid) distance of 10 cm, its weight will be only 16 kg/m$^2$. We can take more cross-section of rod (1.5×3 cm) but more step (up to 15 cm). The weight of grille will be 18 kg/m$^2$. The Dome cover can support 200 – 300 kg/m$^2$ and we can take more strong grilles and armed them by strong artificial fibers.

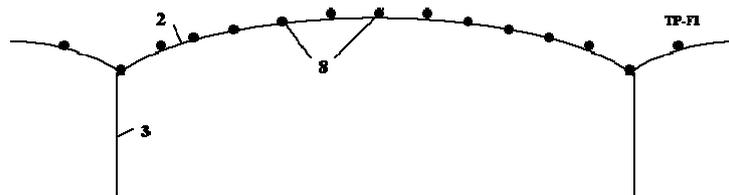

**Fig. 2**. The top film is armored by basalt grille or bonded stones (possibly inside cloth retention pockets) or suspended basalt mesh 8.

**War time**. The offered defense provides protection in the following way. The Qassam or other projectiles have a minimum diameter of 10-17 cm. They, by virtue of their minimum size, must strike the stones or basalt grille, are kinetically destroyed or their trigger is detonated from the shock. The city benefits from the saved lives of its' residents, and gets minimal damage in comparison with what might have occurred. (Even this is greatly reduced if a second optional mesh catches the fragments, which might rain down from an explosion at height. This mesh can be far less sturdy and far tighter than the detonation mesh of steel or basalt above; the object of this second layer, perhaps of Kevlar net, is to catch raining shrapnel as small as a few millimeters. (Even though mostly mesh, this layer can still provide lift by having thin film under parts of the mesh, as little as 10% in a criss-cross pattern; this provides a barrier to the upwelling air aimed at the higher dome and it balloons upwards. These strips should be at least tens of meters across to afford a sizeable stagnation front to the upwelling air.))

An optional third level of coverage would catch down-sifting dust such as anthrax spores from bomblets designed (in the inevitable measure-countermeasure wars) to pierce the outer layer and explode in the breach.

In **peace time** the offered dome produce a fine warm climate (weather) in a covered city. Certain layers of the AB-Dome's films can have a controlled transparency/reflectivity. That allows provision of different solar heating conditions in the city. These gigantic covers are composed of a cheap film, ideally having liquid crystal and conducting layers. The clarity of such may be controlled by electric voltage. They can pass or blockade the solar light (or parts of solar spectrum) and pass or blockade the Earth radiation. The outer and inside radiations have different wavelengths. That makes it possible to control of them separately and to control heating into (and re-radiation from) the Earth's surface. In conventional conditions about 50% of the solar energy reaches the Earth surface. The most part is reflected back to outer space by the white clouds.

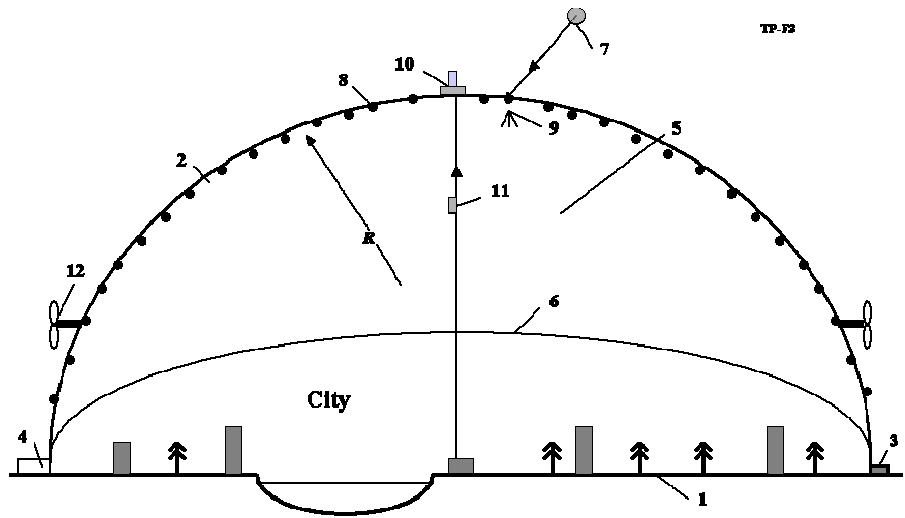

**Fig.3.** Spherical AB-Dome which may provide defense for a given city against rockets, missiles, aviation, chemical and biological weapons. *Notations*: 1 – protected area; 2 – thin film; 3 – ventilator (air pump); 4 – exit; 5 –spherical thin film AB-Dome; 6 – controlled reflectivity thin film (optional) fragment catcher inner layer (optional); 7 –incoming warhead, rocket, missile, bomb or weapon; 8 – strong grille supported by dome; 9 – fragments of destroyed warhead, rocket, missile or aircraft; 10 – TV, communication, telescope, locator emplacements, tourist observation deck; 11 – elevator; 12 – windmills.

In our closed system the clouds (and rain, or at least condensation based dripping) will occur at night when the temperature is low. That means that many cold regions (Alaska, Greenland, Siberia, Northern Antarctica) may absorb more net solar energy and become, within the bubbles, lands with a temperate or sub-tropic climate. That also means the Sahara desert (or locally in Israel, the Dead Sea, Jordan Valley and Negev regions) can be a prosperous area with a fine climate and with a closed-loop water cycle.

The building of a film dome is very easy. Don't think of the popular delusion of a science fiction dome city made of impervious thick crystal, which would be a huge construction project. We simply spread out the film over Earth's surface, turn on the pumping fans and the film is raised by air overpressures to needed altitudes, limited by the support cables.

In case of obstacles in the terrain of the city itself, large areas of film can be readied on a relatively calm day under cable tension at one end of a city, on the edge in the fields, levitated by an overpressure under that film, then pulled across the city, high over head, by towing cables, then locked secure on the far side before final inflation.

Damage to the film is not a major trouble because the additional air overpressure is very small and propeller pumps compensate for any air leakage. Unlike in a space colony or planetary colony, the outside air is friendly and at worst we lose some heat (or cold) and water vapor.

The other advantages of the suggested method include the possibility to paint pictures on the sky (AB-Dome), to show films on the sky by projector, to suspend illuminations, decorations, and air tramways and any other utilities and conveniences (and engineering works) from this new overroof.

Long distance aircraft fly at altitude 8 – 11 kilometers and our AB-Dome (1- 5 kilometers or less) does not trouble them unless the AB-Dome is built on the edge of a glide path to an airport! The support cables will have safety illumination lights (red, flashing, in a string) and internal helicopters will take normal precautions in avoiding contact with them.

More details on the offered AB-Dome are described in [1]-[4], [13]. Additional information is repeated below.

Our design for the AB-Dome is presented in Figures 1-3, which illustrates the thin inflated film dome. The **innovations** are listed here: (1) the construction is air-inflatable; (2) each AB-Dome is fabricated with very thin, transparent film (thickness is 0.05 to 0.3 mm, implying under 150-500 tons a square kilometer) having the control clarity quality without rigid supports; (3) the enclosing film can have (optionally) two conductive layers plus a liquid crystal layer between them which changes its clarity, color and reflectivity under an electric voltage; (4) the boundary section of the AB-Dome has a rounded form. The air pressure is more in these sections and they protect the central sections from the outer wind.

Figures 1-3 illustrate the thin transparent control dome cover we envision. The inflated textile shell—technical "textiles" can be woven or non-woven (films)—embodies the innovations listed: (1) the film is very thin, approximately 0.1 to 0.3 mm. A film this thin has never before been used in a major building; (2) the film has two strong nets, with a mesh of about $0.1 \times 0.1$ m and $a = 1 \times 1$ m, the threads are about 0.5 mm for a small mesh and about 1 mm for a big mesh. The net prevents the watertight and airtight film covering from being damaged by vibration; (3) the film incorporates a tiny electrically conductive wire net with a mesh about 0.1 x 0.1 m and a line width of about 100 $\mu$ (microns, thousandths of a millimeter) and a thickness near 10 $\mu$ (microns). The wire net is an electric (voltage) conductor. It can inform the dome supervisors concerning the place and size of film damage (tears, rips, etc.); (4) the film may be twin-layered with the gap — $c = 1$ m and $b = 2$ m—between covering's layers for heat saving. In polar regions this multi-layered low height covering is the main means for heat insulation and puncture of one of the layers won't cause a loss of shape because the film's second layer is unaffected by holing; (5) the airspace in the AB-Dome's covering can be partitioned, either hermetically or not; and (6) part of the covering can have a very thin shiny aluminum coating that is about $1\mu$ (micron) thick for reflection of unnecessary solar radiation in equatorial or polar regions (without the liquid crystal layer, this aluminizing option is definitely recommended for Israeli use.—it can be preferentially deposited on sections that will receive insolation at higher summer angles, but not on those that will receive it at lower winter angles, just as old-fashioned awnings used to do to keep apartment windows in an acceptable heating range. In case of a near nuclear airburst, this would help reflect damaging thermal rays.) [1]-[4].

## Theory and computation (estimation) of AB-Dome

### a) General information

Our AB-Dome cover (film) has 2 layers (figs. 1,3): top transparent layer 2, located at a maximum altitude (up to 4 –15 kilometers), and lower transparent layer 4 having controllable reflectivity, located at altitude of 1-3 kilometers (optional). The upper transparent cover has a thickness of about 0.05 – 0.3 mm and supports a network of pockets connected by containing hard stones (pebbles, shards of granite). 8. The stones have a mass 0.2 – 1 kg and located about every 0.5 m (step distance).

**Brief information about the cover film.** If we want to control temperature in city, the top film must have some layers: transparent dielectric layer, conducting layer (about 1 - 3 $\mu$), liquid crystal layer (about 10 - 100 $\mu$), conducting layer (for example, $SnO_2$), and transparent dielectric (insulator) layer. Common thickness is 0.05 - 0.5 mm. Control voltage is 5 - 10 V. This film may be produced by industry relatively cheaply per unit area, given the quantities needed.

The conventional controlled clarity (transparency) film reflects superfluous energy back to space. If the film has solar cells integral to it, which will become cheaper as the years go on, then the extra solar energy may be partially converted into electricity.

1. **Liquid crystals** (LC) are substances that exhibit a phase of matter that has properties between those of a conventional liquid, and those of a solid crystal.
  Liquid crystals find wide use in liquid crystal displays (LCD), which rely on the optical properties of certain liquid crystalline molecules in the presence or absence of an electric field. The electric field can be used to make a pixel switch between clear or dark on command. Color LCD systems use the same technique, with color filters used to generate red, green, and blue pixels. Similar principles can be used to make other liquid crystal based optical devices. Liquid crystal in fluid form is used to detect electrically generated hot spots for failure analysis in the semiconductor industry. Liquid crystal memory units with extensive capacity were used in Space Shuttle navigation equipment. It is also worth noting that many common fluids are in fact liquid crystals. Soap, for instance, is a liquid crystal, and forms a variety of LC phases depending on its concentration in water.

   2. **Transparency.** In optics, transparency is the material property of allowing light to pass through.  Though transparency usually refers to visible light in common usage, it may correctly be used to refer to any type of radiation. Examples of transparent materials are air and some other gases, liquids such as water, most glasses, and plastics such as Perspex and Pyrex. Where the degree of transparency varies according to the wavelength of the light. From electrodynamics it results that only a vacuum is really transparent in the strict meaning, any matter has a certain absorption for electromagnetic waves. There are transparent glass walls that can be made opaque by the application of an electric charge, a technology known as electrochromics.Certain crystals are transparent because there are straight lines through the crystal structure. Light passes unobstructed along these lines. There is a complicated theory "predicting" (calculating) absorption and its spectral dependence of different materials.

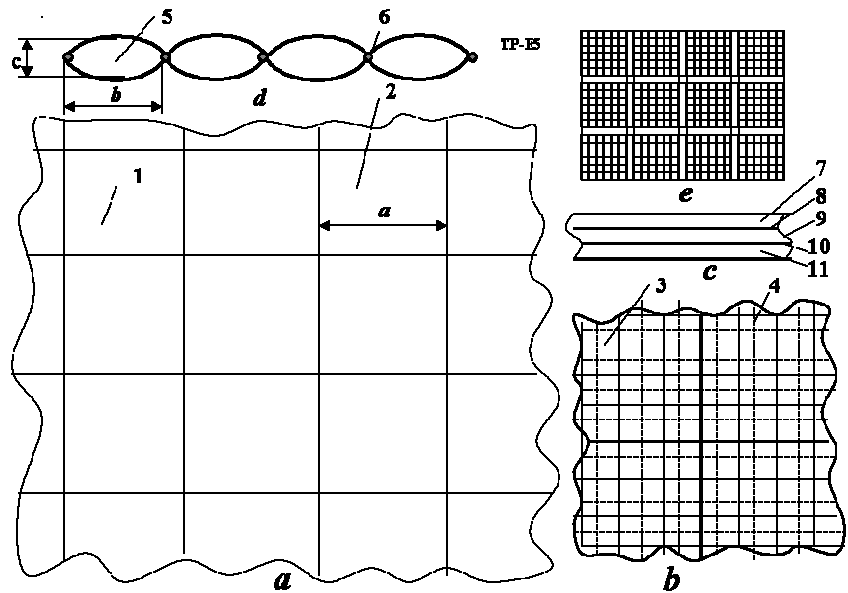

**Fig. 4.** Design of covering membrane. *Notations*: (*a*) Large fragment of cover with controllable clarity (reflectivity, carrying capacity) and heat conductivity; (*b*) Small fragment of cover; (*c*) Cross-section of cover (film) having 5 layers; (*d*) Longitudinal cross-section of low height cover for cold and hot regions (optional); (*e*) Protection grilles. 1 - cover; 2 -mesh; 3 - small mesh; 4 - thin electric net; 5 - cell of cover; 6 - tubes; 7 - transparent dielectric layer, 8 - conducting layer (about 1 - 3 μ (micron)), 9 - liquid crystal layer (about 10 - 100 μ), 10 - conducting layer, and 11 - transparent dielectric layer. Common thickness is 0.1 - 0.5 mm. Control voltage is 5 - 10 V.

   **3. Electrochromism** is the phenomenon displayed by some chemical species of reversibly changing color when a burst of charge is applied.
  One good example of an electrochromic material is polyaniline which can be formed either by the electrochemical or chemical oxidation of aniline. If an electrode is immersed in hydrochloric acid which contains a small concentration of aniline, than a film of polyaniline can be grown on the electrode. Depending on the redox state, polyaniline can either be pale yellow or dark green/black. Other electrochromic materials that have found technological application include the viologens and polyoxotungstates. Other electrochromic materials include tungsten oxide ($WO_3$), which is the main chemical used in the production of electrochromic windows or smart windows.

As the color change is persistent and energy need only be applied to effect a change, electrochromic materials are used to control the amount of light and heat allowed to pass through windows ("smart windows"), and has also been applied in the automobile industry to automatically tint rear-view mirrors in various lighting conditions. Viologen is used in conjunction with titanium dioxide (TiO2) in the creation of small digital displays. It is hoped that these will replace LCDs as the viologen (which is typically dark blue) has a high contrast to the bright color of the titanium white, therefore providing a high visibility of the display.

**4. Film and cable properties** [16]-[19]. Artificial fibers are currently being manufactured, which have tensile strengths of 3-5 times more than steel and densities 4-5 times less than steel. There are also experimental fibers (whiskers) that have tensile strengths 30-100 times more than steel and densities 2 to 5 times less than steel. For example, in the book [16] p.158 (1989), there is a fiber (whisker) $C_D$, which has a tensile strength of $\sigma = 8000$ kg/mm$^2$ and density (specific gravity) of $\gamma = 3.5$ g/cm$^3$. If we use an estimated strength of 3500 kg/mm$^2$ ($\sigma = 7 \cdot 10^{10}$ N/m$^2$, $\gamma = 3500$ kg/m$^3$), than the ratio is $\gamma/\sigma = 0.1 \times 10^{-6}$ or $\sigma/\gamma = 10 \times 10^6$. Although the described (1989) graphite fibers are strong ($\sigma/\gamma = 10 \times 10^6$), they are at least still ten times weaker than theory predicts. A steel fiber has a tensile strength of 5000 MPA (500 kg/sq.mm), the theoretical limit is 22,000 MPA (2200 kg/mm$^2$) (1987); the polyethylene fiber has a tensile strength 20,000 MPA with a theoretical limit of 35,000 MPA (1987). The very high tensile strength is due to its nanotube structure [19].

Apart from unique electronic properties, the mechanical behavior of nanotubes also has provided interest because nanotubes are seen as the ultimate carbon fiber, which can be used as reinforcements in advanced composite technology. Early theoretical work and recent experiments on individual nanotubes (mostly MWNT's, Multi Wall Nano Tubes) have confirmed that nanotubes are one of the stiffest materials ever made. Whereas carbon-carbon covalent bonds are one of the strongest in nature, a structure based on a perfect arrangement of these bonds oriented along the axis of nanotubes would produce an exceedingly strong material. Traditional carbon fibers show high strength and stiffness, but fall far short of the theoretical, in-plane strength of graphite layers by an order of magnitude. Nanotubes come close to being the best fiber that can be made from graphite.

For example, whiskers of Carbon nanotube (CNT) material have a tensile strength of 200 Giga-Pascals and a Young's modulus over 1 Tera Pascals (1999). The theory predicts 1 Tera Pascals and a Young's modules of 1-5 Tera Pascals. The hollow structure of nanotubes makes them very light (the specific density varies from 0.8 g/cc for SWNT's (Single Wall Nano Tubes) up to 1.8 g/cc for MWNT's, compared to 2.26 g/cc for graphite or 7.8 g/cc for steel). Tensile strength of MWNT's nanotubes reaches 150 GPa.

Specific strength (strength/density) is important in the design of the systems presented in this paper; nanotubes have values at least 2 orders of magnitude greater than steel. Traditional carbon fibers have a specific strength 40 times that of steel. Since nanotubes are made of graphitic carbon, they have good resistance to chemical attack and have high thermal stability. Oxidation studies have shown that the onset of oxidation shifts by about 100$^0$ C or higher in nanotubes compared to high modulus graphite fibers. In a vacuum, or reducing atmosphere, nanotube structures will be stable to any practical service temperature (in vacuum up 2800 $^o$C. in air up 750$^o$C).

In theory, metallic nanotubes can have an electric current density (along axis) more than 1,000 times greater than metals such as silver and copper. Nanotubes have excellent heat conductivity along axis up 6000 W/m·K. Copper, by contrast, has only 385 W/m·K.

About 60 tons/year of nanotubes are produced now (2007). Price is about $100 - 50,000/kg. Experts predict production of nanotubes on the order of 6000 tons/year and with a price of $1 – 100/kg to 2012.

Commercial artificial fibers are cheap and widely used in tires and countless other applications. The authors have found only older information about textile fiber for inflatable structures (Harris J.T., Advanced Material and Assembly Methods for Inflatable Structures, AIAA, Paper No. 73-448, 1973). This refers to DuPont textile Fiber **B** and Fiber **PRD-49** for tire cord. They are 6 times strong as steel (psi is 400,000 or 312 kg/mm$^2$) with a specific gravity of only 1.5. Minimum available yarn size (denier) is 200, tensile module is 8.8×10⁶ (**B**) and 20×10⁶ (**PRD-49**), and ultimate elongation (percent) is 4 (**B**) and 1.9 (**PRD-49**). Some data are in Table 1.

**Table 1.** Material properties

| Material | Tensile strength kg/mm² | Density g/cm³ | **Fibers** | Tensile strength kg/mm² | Density g/cm³ |
|---|---|---|---|---|---|
| Whiskers | | | | | |
| AlB$_{12}$ | 2650 | 2.6 | QC-8805 | 620 | 1.95 |
| B | 2500 | 2.3 | TM9 | 600 | 1.79 |
| B$_4$C | 2800 | 2.5 | Allien 1 | 580 | 1.56 |
| TiB$_2$ | 3370 | 4.5 | Allien 2 | 300 | 0.97 |
| SiC | 1380-4140 | 3.22 | Kevlar or Twaron | 362 | 1.44 |
| **Material** | | | Dynecta or Spectra | 230-350 | 0.97 |
| Steel prestressing strands | 186 | 7.8 | Vectran | 283-334 | 0.97 |
| Steel Piano wire | 220-248 | | E-Class | 347 | 2.57 |
| Steel A514 | 76 | 7.8 | S-Class | 471 | 2.48 |
| Aluminum alloy | 45.5 | 2.7 | Basalt fiber | 484 | 2.7 |
| Titanium alloy | 90 | 4.51 | Carbon fiber | 565 | 1,75 |
| Polypropylene | 2-8 | 0.91 | Carbon nanotubes | 6200 | 1.34 |

Source: [15]-[18]. Howatsom A.N., Engineering Tables and Data, p.41.

Industrial fibers have up to $\sigma = 500\text{-}600$ kg/mm², $\gamma = -1800$ kg/m³, and $\sigma/\gamma = 2{,}78 \times 10^6$. But we are projecting use in the present projects the cheapest films and cables applicable (safety $\sigma = 50 - 100$ kg/mm²).

**5. Wind effect.** As wind flows over and around a fully exposed, nearly completely sealed inflated AB-Dome, the weather affecting the external film on the windward side must endure positive air pressures as the wind stagnates. Simultaneously, low air pressure eddies will be present on the leeward side of the AB-Dome. In other words, air pressure gradients caused by air density differences on different parts of the AB-Dome's envelope is characterized as the "buoyancy effect". The buoyancy effect will be greatest during the coldest weather when the AB-Dome is heated and the temperature difference between its interior and exterior are greatest. In extremely cold climates such as the Arctic and Antarctic Regions the buoyancy effect tends to dominate AB-Dome pressurization.

**6. Solar radiation.** Solar radiation impinging the orbiting Earth is approximately 1400 W/m². The average Earth reflection by clouds and the sub-aerial surfaces (water, ice and land) is about 0.3. The Earth-atmosphere absorbs about 0.2 of the Sun's radiation. That means about $q_0 = 700$ W/m²s of solar energy (heat) reaches our planet's surface in cloudy weather at the Equator. That means we can absorb about 30 - 80% of solar energy. It is enough for normal plant growth in wintertime (up to 40-50° latitude) and in circumpolar regions with a special variant of the AB-Dome design.

The solar spectrum is graphically portrayed in Fig. 5.

The visible part of the Sun's spectrum is only $\lambda = 400 - 800$ nm (0.4 to 0.8 $\mu$.). Any warm body emits radiation. The emission wavelength depends on the body's temperature. The wavelength of the maximum intensity (see Fig. 5) is governed by the black-body law originated by Max Planck (1858-1947):

$$\lambda_m = \frac{2.9}{T}, \quad [mm], \qquad (1)$$

where $T$ is body temperature, °K. For example, if a body has an ideal temperature ~20 °C ($T = 293$ °K), the wavelength is $\lambda_m = 9.9$ μ.

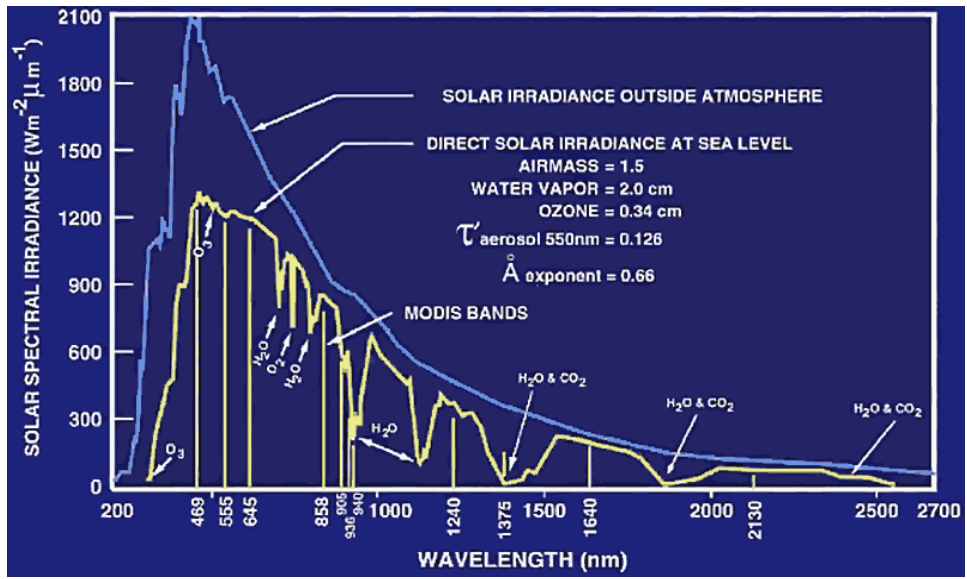

**Fig. 5**. Spectrum of solar irradiance outside atmosphere and at sea level with absorption of electromagnetic waves by atmospheric gases. Visible light is 0.4 - 0.8 μ (400 – 800 nm).

The radiation energy emitted by a body may be computed by employment of the Josef Stefan-Ludwig Boltzmann law:

$$E = \varepsilon \sigma_s T^4, \quad [W/m^2], \tag{2}$$

where $\varepsilon$ is coefficient of body blackness ($\varepsilon = 0.03 \div 0.99$ for real bodies), $\sigma_s = 5.67 \times 10^{-8}$ [W/m²·K] Stefan-Boltzmann constant. For *example*, the absolute black-body ($\varepsilon = 1$) emits (at $T = 293\ ^0K$) the energy $E = 418$ W/m².

**7. Earth's atmosphere**. The property of Earth's atmosphere needed for computations are presented in Table 2 below.

**Table 2.** Standard Earth atmosphere

| $H$ km | 0 | 1 | 2 | 3 | 4 | 5 | 6 | 7 |
|---|---|---|---|---|---|---|---|---|
| $\bar{p} = p_h/p_o$ | 1 | 0.887 | 0.784 | 0.692 | 0.609 | 0.533 | 0.466 | 0.406 |
| $H$ km | 8 | 9 | 10 | 11 | 12 | 13 | 14 | 15 |
| $\bar{p} = p_h/p_o$ | 0.362 | 0.304 | 0.261 | 0.224 | 0.191 | 0.164 | 0.14 | 0.12 |

**4. The thickness and weight of the AB-Dome envelope**, its sheltering shell of film, is computed by formulas (from equation for tensile strength):

$$\delta_1 = \frac{Rp}{2\sigma}, \quad \delta_2 = \frac{Rp}{\sigma}, \tag{3}$$

where $\delta_1$ is the film thickness for a spherical dome, m; $\delta_2$ is the film thickness for a cylindrical dome, m; $R$ is radius of dome or radius of cover cell between cable (it may be half of distance between top cable, m; $p$ is additional pressure into the dome, N/m², ($p$ depends from altitude); $\sigma$ is safety tensile stress of film, N/m².

For *example*, compute the film thickness for a dome having radius $R = 100$ m (distance between top cable 7 is 400 m), additional air pressure $p = 0.01$ atmosphere ($p = 1000$ N/m²), safety tensile stress $\sigma = 50$ kg/mm² ($\sigma = 5 \times 10^8$ N/m²), spherical dome. We receive $\delta_1 = 0.1$ mm. Distance between main cable 3 is $D = 0.8$ kilometers (Fig. 6).

The computation for others case are presented in Fig. 7 below.

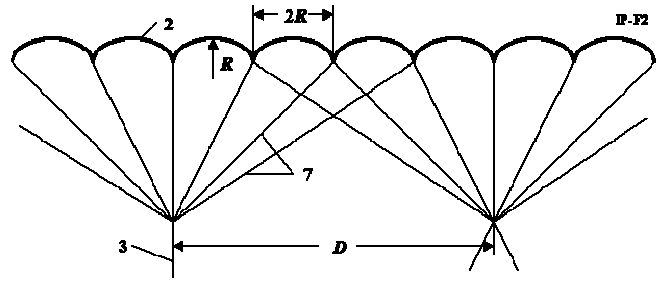

**Fig. 6**. Cable support system. Radius $R$ spherical cell of dome cover and distance $D$ between main cable.

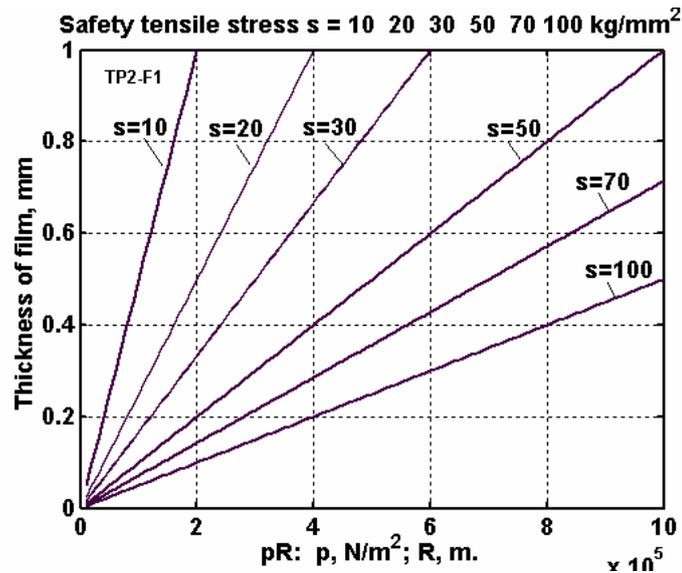

**Fig.7.** The thickness of top cover via the production of overpressure and radius spherical dome cell (distance between top cables for different safety film tensile stress.

The cover weight (mass) of 1 m² is computed by the equation:
$$m = \gamma \delta, \qquad (4)$$
where $m$ is 1 m² film mass, kg/m²; $\gamma$ is cover density, m. For *example*, if the cover thickness is $\delta = 0.2$ mm $= 0.0002$ m and $\gamma = 1500$ kg/m³, the $m = 0.3$ kg/m².

Area $S_c$ of semi-sphere diameter $R$, film cover mass $M_f$ and cost $C$ of Dome cover are
$$S_c = 2\pi R^2, \quad M_f = m_f S_c, \quad C = cS_c, \quad C = c_m M_f \ , \qquad (5)$$
where $R$ is radius of semi-sphere, m; $m_f$ is average cover area of 1 m² ; $c$ is cost of 1 m², $US/m²; $c_m$ is cover cost of 1 kg, $US/m²; $C$ is cost of total cover, $US.

*Example*. Let us take the hemi-sphere dome case. If $m_f = 0.3$ kg/m², film cost $c = \$0.1$ /m². The film mass covered of 1 km² of ground area is $M_1 = 2 \times 10^6$ $m_c = 600$ tons/km² and film cost is \$60,000/km². Fabrication costs may be somewhat more.

The area of city diameter 2 kilometers is 3.14 km². Area of a 2 kilometer hemi-spherical dome is 6.28 kilometers². The material cost of such an AB-Dome cover is theoretically 0.628 millions $US.

The total cost of installation would be about 3-9 million $US if the Government of Israel would make available free army labor to fabricate/assemble the shell and prepare ground installations.

That is less by nearly a hundred times, than the cost of an anti-rocket system (some hundreds of millions $US), and by over a thousand times for an anti strategic nuclear missile system (billions of $ US) The anti-rocket system is useless in peace time and it may be useless soon in war time because the offensive weapon is easily improved (multiple warheads, dispersing lethal substances, change of attack profile) . The offered AB-Dome is very useful in peace time (controlling the weather and temperature inside, which could enable saving the AB-Dome cost in a few years of heating or air conditioning savings.), The AB-Dome defense may be upgraded to be proof also against any likely biological, chemical, radioactive dispersal weapons. The AB-Dome is a CLOSED-LOOP system (exclusive of leaks, but even 99% closure reduces outer-world threats by a hundred fold, and in practice even greater ratios should be obtainable).

The most extreme cases are deeply unpleasant, but must be discussed in the current Israeli threat environment. An incoming 20-kiloton missile warhead could be stopped by an AB-Dome and either impact-killed (relatively harmlessly) or detonated at a height too great to destroy the city below. In the latter case, assuming a 5 kilometers burst height and aluminized (summer sun inclination) upper dome sections, obviously the upper portions of the dome would be vaporized directly around the warhead, but the further sections would survive with progressively less damage, especially if the dome itself were sectioned internally (recommended). Because the AB-Dome contains positive overpressure internally, radioactive dust particles from the bomb or casing would tend to drift by even leaking sections of the AB-Dome outside the destroyed section. And the city below in the 'nuked' quarter would still be repairable and nearly uncontaminated by the standards of a low-altitude detonation (unstopped by the AB-Dome), which burst at optimal damage inflicting height (Mach effect enhanced).

Even more severe damage would have resulted from a surface detonation, which could easily have left a nearly 100 meter crater complex, which itself could cost more to decontaminate than the cost of the AB-Dome if there were *no* damage in the shielded city! (As well as unknown contamination plumes in groundwater and aquifers downstream)

(In addition, at 10 kilometers distance from a ground burst radiation from fallout would give an unshielded person a lethal dose *within 1 hour.* And crops from that land might be unsalable for generations. If a ground burst can be avoided, all these damages are greatly mitigated. (see for example http://www.epp.cmu.edu/domesticsecurity/KeithFlorig.pdf)

Consider the contrast in refugee problems: With undomed cities the entire small city is effectively rendered uninhabitable by an attack, (the Chernobyl-like panic about spreading radiation alone would see to that) and ironically, those fleeing the first blast may inadvertently head straight into a second target's fatal fallout plume!

By contrast, with a 20 kiloton explosion at 5 kilometers altitude impacting and detonating on the outer apex barrier of a 5 km high AB-Dome, winds at the surface underneath point zero (assuming they penetrated any optional lower barrier) would be no higher than 21 meters per second, or about 47 miles per hour. Because of resonance effects, it is quite possible that every window in the AB-Dome compartment under the detonation might be broken, but the much mocked 'duck and cover' training would literally be enough to avoid significant casualties (from heat pulse and flying glass). People could stay in their homes, sweep up, and await instructions. (In the compartment of the AB-Dome directly under the detonation, those instructions would presumably be to evacuate promptly to an adjoining compartment until the dome section could be repaired, because the city *in that compartment* is now open to a ground-level blast.

But the tens of thousands of dead, the unrecognizable charred bodies, the mass graves and all the other phenomena that Israelis are already quite familiar with from the Holocaust would not have taken place.

Obviously this is no magic solution or panacea for incoming nuclear missiles, (especially against ground-smuggled weapons—although the AB-Dome's restricted and monitored entrance points would aid in that other struggle).

People worry if any defense against missiles at all is developed, it will destabilize the concept of deterrence (or ignite an arms race) but we think a calm reading of the capabilities described would lead to the certain conclusion that only a crazy person would confuse 'a second chance at life' with 'immunity against all attacks'. To lend perspective to the matter, during the Cold War, a single US Titan II or a Soviet SS-9 could easily have destroyed the entire 5 km AB-Dome and underlying small city with a single shot. The lethal radius is the size of the entire AB-Dome! (This is one theoretically interesting case where the cubic kill capacity of a very large thermonuclear weapon would not be wasted, because the dome would be a target in three dimensions, not just two.)

To clarify: *This is a partial, final layer defense* only *against low-yield weapons launched in anger or accident by novice nuclear powers with unstable command and control networks.*

However, with India already with a hydrogen bomb, some Pakistanis calling to match it, China so armed now and the Japanese studying the situation, once can foresee at some point in the future a crazy world of the type Sir Martin Rees ("Our Final Hour") among others worries about; where apocalyptic war of some sort takes place between various super-weaponed regional contestants every other generation. In such a extreme case, like a early 1960's science fiction horror story, much of Earth could be poisoned by radioactive precipitations, tailored poison vapors and gases, genetically engineered killer microbes-- but a city (or even major portions of a country such as the best farmland) could continue to exist uncontaminated under AB-Domes and not only enable national survival and economic recovery, but hopefully be able to render some sort of aid to the survivors of the war—a lifeboat, one of many, on a national scale.

*Disclaimer: Both the authors (and hopefully the reader) find the above scenario horrific and terrifying, and we suppose in today's academic climate it's actually necessary to state openly that we neither look forward to nor endorse crazy whole-population killing attacks or war plans. Just for the record. But given the other record, the historical record of treatment of civilians of the last hundred years, and the scientific prospects for the next hundred, we do endorse having lifeboats on any vessel we sail on, even if those lifeboats never do leave Spaceship Earth.*

**5. The mass of support cable for 1 m² projection of dome cover.** The mass of the support cable for every projection 1 m² of dome cover may be computed by the equation:

$$m_c \approx \gamma_c \frac{\overline{p}p}{\sigma} H, \quad M_c = \sum_S m_c, \quad M_c \approx \gamma_c \frac{p_a}{\sigma} S H_a, \tag{6}$$

where $m_c$ is cable mass supported the projection one m² of the dome cover, kg/m²; $M_c$ is total mass of cable, kg; $S = \pi R^2$ is projection of cover on ground, m²; $\gamma_c$ is density of cable, kg/m³; $p$ is overpressure, N/m²; $\sigma$ is safety cable tensile stress, N/m²; $H$ is height of cable, m; $\overline{p}$ is relative air pressure at given altitude (Table 2); $p_a$ is average overpressure, N/m²; $H_a$ is average height of the support cable, m.

*We can design the dome cover without the support cable.* In this case we compute the thickness of dome cover for radius $R$ of a full Dome (see Eq.(3)). The cover thickness will be more. That is better for defense from rockets. If we want to compute more exactly and spend less cover mass, we must compute the variable overload (from altitude). In this case we must take into account the variable thickness of dome cover. If the dome doesn't have the support cable, one can have suspended cables, which allow reaching any part of Dome. The non-cable AB-Dome requires a (about 3-4 times) thicker cover that increases the impact force of the warhead and may make the stones unneeded.

For increased protection the dome cover may be an armored strong grille. The missile is destroyed from the shock of impact. It is not necessary to apply explosive of ours against it.

**6. Lift force of 1 m² projection of dome cover.** The lift force of 1 m² projection of dome cover

is computed by equation:

$$L_1 = \bar{p}p, \qquad (7)$$

where $L_1$ is lift force of 1 m² projection of dome cover, N/m²; $p$ is overpressure, N/m²; $\bar{p}$ is relative air pressure at a given altitude (Table 2).

*Example*, for $p = 0.01$ atmosphere $= 1000$ N/m², $H = 10$ kilometers ($\bar{p} = 0.261$, Table 2) we get $L_1 = 261$ N $= 26$ kg.

Result of computation for different $p$ is shown in Fig.11.

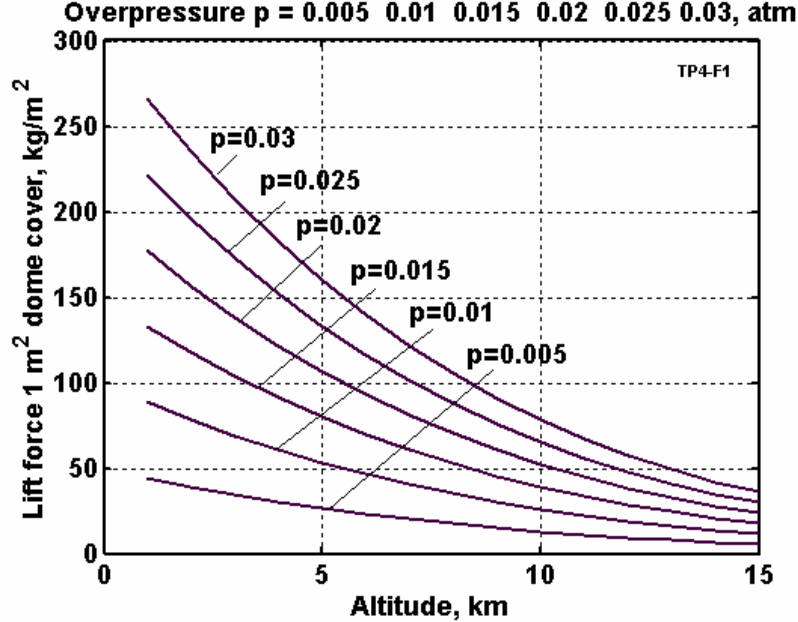

**Fig.8**. Lift force 1 m² of vertical projection the Dome cover versus altitude for different overpressures.

1. **Leakage of air through hole.** The leakage of air through hole, requested power of levitation fans (ventilator), and time of sinking of Dome cover (in case of large hole) may be estimated by the equation:

$$V = \sqrt{\frac{2p}{\rho}}, \quad M_a = \rho V S_h, \quad N = \frac{pVS_h}{\eta}, \qquad (8)$$

where $V$ is speed of air leakage, m/s; $p$ is overpressure, N/m²; $\rho$ is air density at given altitude, $\rho = 1{,}225$ kg/m³ at $H = 0$; $S_h$ is area of hole, m; $N$ is motor power, W; $\eta$ is coefficient efficiency of motor.

*Example.* The area of hole equals $S_h = 10^2$ m² (10×10 m) at $H = 0$ m, $p = 0.001$ atmosphere $= 100$ N/m², $\eta = 0.8$. Computation gives the $V = 12.8$ m/s, $M_a = 1568$ kg/s, $N = 161$ kW.

Let us estimate the time required for the AB-Dome cover to sink if no air is pumping it upward. Take the sphere radius $R = 500$ m. The volume of semi-sphere is $v = 262 \times 10^6$ m³, the air rate is $q = VS_h = 12.8 \times 100 = 1280$ m³. The time of the full sinking is $t = v/q = 204 \cdot 10^3$ s $= 57$ hours.

Note, the overpressure-driven air will flow out from the dome. That means the radioactive (or chemical, biological, or other) bomb pollution cannot penetrate within the AB-Dome. The outer air pumped into the AB-Dome to levitate it can be filtered from radioactive or biological or chemical dust.

The repair of a dome hole is easy; no heroic measures are required at altitude. The support cable of the needed part of dome is reeled in and a film patch closes the hole. In case of a nuclear detonation at altitude this might effectively be most of a new dome section; presumably stockpiling repair materials en masse before a crisis will be a wise move.

**9. Protection grille**. For increased protection the dome cover may be armored with strong grilles (for example, steel rods having diameter 1 cm and step 10 cm), contain a mass 10 – 16 kg/m². All grilles are connected by flexible cables.

**10. The wind dynamic pressure** from wind is:

$$p_w = \frac{\rho V^2}{2}, \qquad (9)$$

where $\rho = 1.225$ kg/m³ is air density; $V$ is wind speed, m/s.

For example, a storm wind with speed $V = 20$ m/s, standard air density is $\rho = 1.225$ kg/m³. Than dynamic pressure is $p_w = 245$ N/m². That is four times less when internal pressure $p = 1000$ N/m² = 0,01 atm. When the need arises, sometimes the internal pressure can be voluntarily decreased, bled off.

# Projects

Let us to consider the protection of a typical small city having diameter of 2 kilometers by semi-spherical AB-Dome, having radius (altitude) $R = 1$ kilometers. The covered region is $S = \pi R^2 = 3.14$ km², the semi-spherical area is $S_s = 2S = 6.28$ km².

The many computations for this AB-Dome are made as examples in the theoretical section above. We summarize these data in one project.

Let us take the overpressure as 0.1 atmosphere at sea level. Note that divers routinely tolerate overpressures of some atmospheres. The Earth's atmosphere changes the pressure by some percentage points locally and we don't feel it. (Take a film having safety tensile stress $\sigma = 100$ kg/mm², local $R = 1000$ m. Dome is taken as a hemi-sphere without the support cables. Then the film has a thickness $\delta_1 = 0.5$ mm. The second lower film may be 0.1 – 0.2 mm thickness.

Area of semi-sphere is $S_s = 2\pi R^2 = 6.28$ km². If density of film equals $\gamma = 1500$ kg/m³, the total mass of cover film is $M = \gamma \delta S_s = 4,700$ tons. If cost of film is $1/kg, the total material cost of top cover is C = $4.7 millions.

We armor the city from one side (area 3,14 km²). The candidate materials are basalt grille (rebar mesh) or large pebble-sized hard stones (say 10 centimeters) suspended in cloth pockets tensioned by straps in a grid. The stones or grille may be armored by basalt fiber. Let us take the grid size of basalt grille as $s = 10$ cm, the rod cross-section 1.5×3 cm and the mass equal to $m_s = 27$ kg/m². The total mass of grilles is $M_s = m_s S_s = 85,000$ tons. If rebar price of the cast basalt (grille) is $100 a ton (January 2008) 85,000 tons would be $8.5M. If the stones' price is $50/ton, the stone costs <$5 millions. The cloth pockets and straps may be $1 million more. Adopting that option, the total cost of material is $10.7 million from stones or $14.2 with basalt grille.

The total mass of construction is 54,700 – 85,000 tons and cost of construction material $10.7 - $14.2 million. The total cost of AB-Dome cover over a small city is about $14 -17 million.

Our computation of this AB-Dome is far from being optimized at present. The average lift force (100 – 300 kg/m²) is over in 4 - 12 times the weight of cover (0.75 kg/m²) plus weight of stones (16 kg/m²) or grille (27 kg/m²). We can increase the mass (and number) of the pocketed stones or basalt impact grille by 3-5 times or decrease the thickness of the film cover.

**Defense of Sederot**

> - It is a low-resolution image, deliberately decolorized, and thus not suitable for production of counterfeit goods.
> - The logo is not used in such a way that a reader would be confused into believing that the article is written or authorized by the owner of the logo.

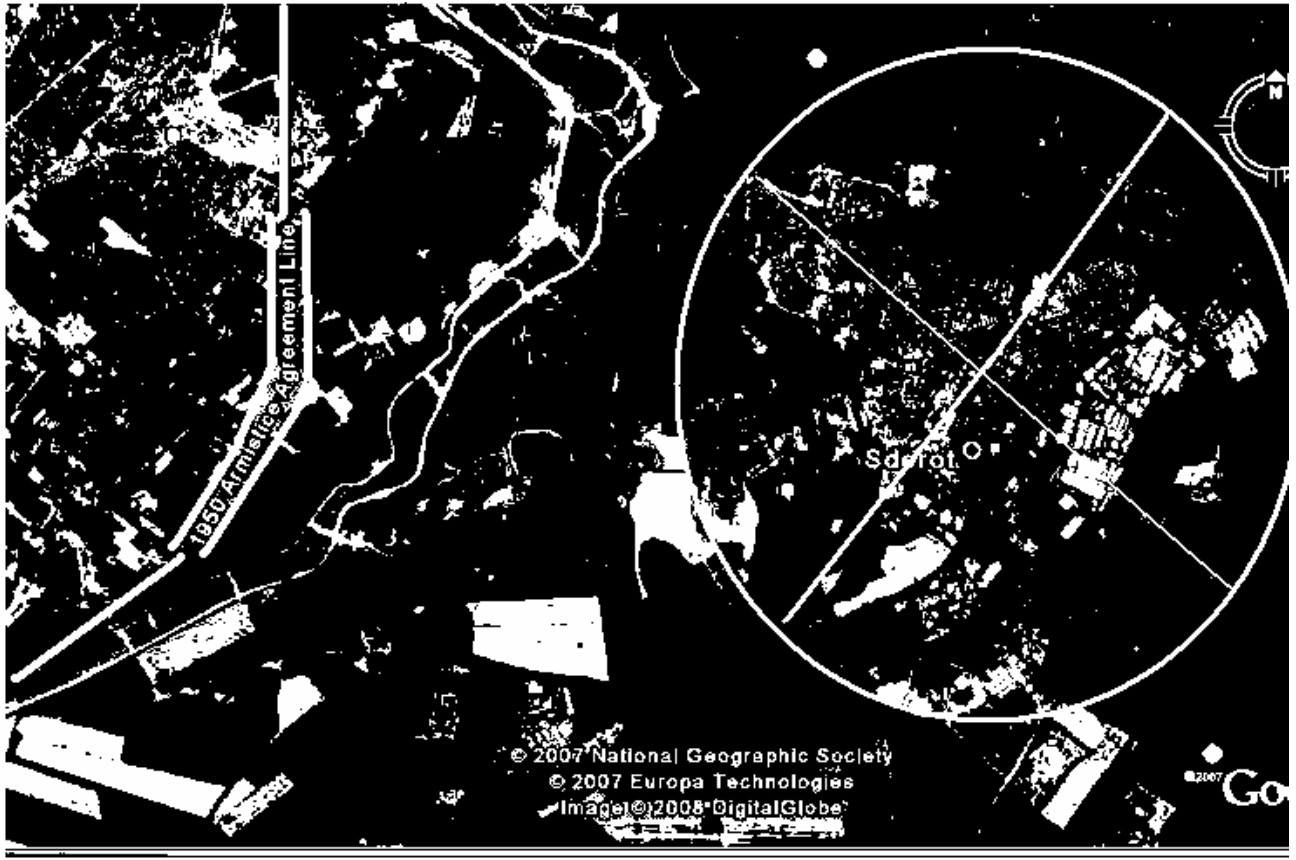

In the above edited Google Earth picture the 1950 Gaza border is seen at left. The dome on the right is 4.5 kilometers on a side; smaller domes might do better for subsections, but we compute very roughly the parameters needed for Sederot's defense.

A 4.5 kilometer wide dome, centered roughly at 31°31'19.75"N, 34°35'52.18"E as outlined above, would cover the widest portion of the built up area, with plenty of agricultural land under the dome as well.
The area relative to the previously computed project is ~5.07 the area and expense.

Thus, 1 layer of cover (dome material) is ~5.07 × 4.7 = 23.829 million $US. (2 layers are recommended for defense against 20-kiloton missile warheads, and for internal segregation)
Total material including stone cover is ~5.07 × 10.7 = 54.249 million $US.
Total cost of AB-Dome cover over Sederot is about ~5.07 * $14 -17 million or $~81.12 million $US. With the 20-kiloton protection option, the cost rises to above $105 million.
Even this is less than half the 1.5 billion NIS (New Israeli Shekels) allocated for the new missile defense system, and would enable an end to alerts, refugees and a city in Israel being held hostage at a cost of incalculable damage to Israel's deterrent power. With the missile defense system, the game simply goes on until the enemy figures out easily achieved new countermeasures. (Active offense nearly always beats out active defense; passive defense usually beats both in civil defense terms)

The authors are prepared to discuss the problems with organizations which are interested in research and development related projects.

## Discussion

   The cost of the offered passive AB-Dome Protection System may be less by hundred of times, then the cost of an *effective* active anti-rocket system (tens of billions $US). The anti-rocket system is useless in peacetime and it may be useless soon in wartime because of easily adopted enemy countermeasures. The offered AB-Dome is very useful in peacetime (especially for controlling the weather and temperature inside, which could enable saving the AB-Dome cost in a few years of heating or air conditioning savings.)

   Additional benefits: Rainwater can run to collection troughs unmixed with sewer water or street pollution directly from the AB-Dome. This can be a few hundred thousand tons for a couple of square kilometers per year—of very high purity. This is no small matter in a rain-short country like Israel. Also, farms within will have evaporation of water from the fields during day—and condensation and collection at night—a closed cycle system, the water 'capital' subtracted from only by final export of the grown produce—a tiny percentage of the life cycle water intake of the crops. (For example a steer might drink 24 tons of water in its lifetime, be washed with 7, and be fed crops containing 5221 tons of water—but only the steer's final weight must leave the agricultural dome. Data is taken as to potential water savings (from **http://www.waterfootprint.org/**-- 900 liters of water to grow 1 kilogram of corn (maize); 16000 liters of water to grow 1 kg of beef. Instead of being a sign of waste, this would be the measure of conservation instead! Respectively, .9 tons and 16 tons of water would remain in the dome—1 kg of packaged product in each case would leave it.)

Besides this vast water savings (enabling desalination to actually be economical to grow crops in the present day (though recycling water, and only making up the net system leaks) tropical crops could be grown in temperate countries with such domes. Extended growing seasons through heat containment would be possible in most locations, and multiple growing seasons in many, doubling the productivity of the land. And the crops would be protected against external migrating insects such as locusts, crop and bird diseases, migrating pests or wildlife that eat crops, frosts, and being downwind of nuclear, biological or chemical contamination. (In the year of Chernobyl many Eastern European countries had great difficulty marketing even their *uncontaminated* crops to overseas buyers—people simply did not wish to take any chances, no matter how small, as we see in the present GM (genetically modified) crops controversy.
 —And incidentally, non-GM crops, or 100% GM crops, could be isolated within subsections of an AB-Dome, and be able to prove it to their target markets.) It is no exaggeration to say that in the futures farmers wanting bank financing will have to prove they are under a dome system *to make the odds of growing and marketing their crops successfully good enough to bank on.*
 The AB-Dome City Defense System also provides protection from nearly all external terrorist threats imaginable, even a 9-11 type airliner incident, even a dirty bomb upwind (or fallout from a nuclear war around the world, would be kept literally kilometers away from the center of the city. Given the strong financial, water conservation and security incentives to dome over good croplands (and dome over cities, as well) –doing both would together provide a powerful degree of national continuity in an environmentally and security challenged place like Israel.

   The control of regional and global weather of the Earth is an important problem of humanity. That ability dramatically increases the territory suitable for men to live in, the sown area, and crop capacity. In the long term, it allows us to convert all Earth land such as Alaska, North Canada, Siberia, deserts Sahara or Gobi to a prosperous garden. The suggested method is very cheap (cost of covering 1 m$^2$ is about 2 - 15 cents) and may be utilized at the present time. We can start from a small area, from small towns in bad climactic regions and extended to a large area.

Film domes can foster the fuller economic development of cold regions such as the Earth's Arctic and Antarctic and, thus, increase the effective area of territory dominated by humans. Normal human health can be maintained by ingestion of locally grown fresh vegetables and healthful "outdoor" exercise. The domes can also be used in the Tropics and Temperate Zone. Eventually, they may find application on the Moon or Mars since a vertical variant, inflatable towers to outer space, are soon to become available for launching spacecraft inexpensively into Earth-orbit or interplanetary flights. An AB-Dome can keep at high altitude a load up to 300 kg/sq.m. A launch site on such a lofted dome would have the same advantage as from a mountaintop, yet be buildable, for example, at the ESA and NASA tropical launch sites, easily accessible by cable car from their existing surface bases. Because of the exponential nature of the rocket equation, by taking off from 5 km elevation, even a minor thrust augmentation because of the thinner atmosphere would probably enable enough extra fuel being loaded to nearly double a given launcher's payload to orbit. (When one additionally takes into account the reduced drag losses)

Lest it be objected that such domes as considered in this article would take impractical amounts of plastic, consider that the world's plastic production is today on the order of 100 million tons. If, with economic growth, this amount doubles over the next generation and the increase is used for doming over territory, at 300-500 tons a square kilometer 200,000 square kilometers could be roofed over annually. While small in comparison to the approximately 150 million square kilometers of land area, consider that 200,000 one kilometer sites scattered over the face of the Earth newly made productive and more habitable could revitalize vast swaths of land surrounding them—one square kilometer here could grow local vegetables for a city in the desert, one over there could grow biofuel, enabling a desolate South Atlantic island to become fuel independent; at first, easily a billion people a year could be taken out of sweltering heat, biting cold and slashing rains, saving the money buying and running heating and air conditioning equipment would require. Additionally, clean rain water could flow directly to cisterns, away from the pollution of the storm sewers. And if needed, bio-oil crops could be grown within to enable further dome production—*possibly using carbon dioxide exhaust from enclosed power stations* to hasten growth rates.

In effect, by doming over inhospitable land as specified, in exchange we get new territory for living with a wonderful climate, largely free from serious attacks, even if right near a hostile border.

The associated problems are researched in references [1]-[12].

## Results

Authors offer the cheap AB-Dome which protects the big cities from rockets, mortar shell, projectiles, chemical, biological weapon (bombs) delivered by rockets, missiles, guns and aviation. The offered AB-Dome is also very useful in peacetime because that protects the city from ambient weather and creates a very livable climate inside the Dome.

The principal advantages of the offered AB-Dome follow:
1. AB-Dome may be cheaper by hundreds to thousands of times than current comparable anti-rocket systems.
2. AB-Dome may not need high technology and can built by a poor country.
3. Additional yearly crops, high-yield growing conditions, conservation of scarce water, and freedom from pesticides, genetic drift and damaging factors for any crops grown within. (This also may enable its financing).
4. By making useless land valuable, the dome may be able to enable its own financing.
5. It is easy to build.
6. AB-Dome may be used in peacetime; it creates a fine climate (weather) inside the dome and thus can be sold to a national assembly on air conditioning and heating savings alone.
7. AB-Dome may protect from rocket, mortar, chemical, biological and worse weapons.

8. Dome protects the autonomous existence of the city population after third-party nuclear or biological war and spreading contamination of the regional atmosphere.
9. AB-Dome may be used for high antennae site for regional TV, for cell phone communication, for long distance location (differential GPS service), for astronomy (telescope above ~50% of air thickness). (This also may enable its financing).
10. Unlike any known active ballistic missile defense, this AB-Dome also can defend against submunitions and dispersed killing agents such as shrapnel.
11. AB-Dome may be used for high altitude tourism, or support for high altitude activities.
12. AB-Dome may be used for the high altitude windmills (getting of cheap renewable wind energy).
13. AB-Dome may be used for night illumination and entertainment or advertising (this also may enable its financing).

Additional applications of the offered AB-Dome the reader may find in [1]-[15].

## REFERENCES

(The reader may find some of these articles at the author Bolonkin's web page: http://Bolonkin.narod.ru/p65.htm, http://arxiv.org , search "Bolonkin", in the book "*Non-Rocket Space Launch and Flight*", Elsevier, London, 2006, 488 pgs. and in book "New Concepts, Ideas, Innovations in Aerospace, Technology and Human Science", NOVA, 2008, 430 pgs.)